\documentclass[letterpaper,11pt]{article}
\pdfoutput=1 

\usepackage{jheppub} 

\usepackage{dsfont}

\def\ben{\begin{equation}}
\def\een{\end{equation}}

\title{Three Dimensional View of the SYK/AdS Duality}

\author{Sumit R. Das$^1$,}
\author{Antal Jevicki$^2$,}
\author{Kenta Suzuki$^2$,}

\affiliation{$^1$Department of Physics and Astronomy, University of Kentucky, Lexington, KY 40506, U.S.A.}
\affiliation{$^2$Department of Physics, Brown University, 182 Hope Street, Providence, RI 02912, U.S.A.}

\emailAdd{das@pa.uky.edu}\emailAdd{antal\_jevicki@brown.edu}\emailAdd{kenta\_suzuki@brown.edu}

\abstract{We show that the spectrum of the SYK model can be interpreted as that of a 3D scalar coupled to gravity.
The scalar has a mass which is at the Breitenholer-Freedman bound of AdS$_2$, and subject to a delta function potential at the center of the interval along the third direction.
This, through Kaluza-Klein procedure on AdS$_2 \times (S^1)/Z_2$, generates the spectrum reproducing the bi-local propagator at strong coupling.
Furthermore, the leading $1/J$ correction calculated in this picture reproduces the known correction to the poles of the SYK propagator,
providing credence to a conjecture that the bulk dual of this model can be interpreted as a three dimensional theory.}

\begin{document}

\begin{flushright}
{UK/17-03} \\ {BROWN-HET-1713}
\end{flushright}

\maketitle
\flushbottom

\section{Introduction}
\label{sec:intro}
Recently, the Sachdev-Ye-Kitaev (SYK) model \cite{Kitaev:2015, Kitaev:2014, Sachdev:2015efa, Polchinski:2016xgd, Maldacena:2016hyu, Jevicki:2016bwu, Jevicki:2016ito, Davison:2016ngz},
which arose from Sachdev-Ye model \cite{Sachdev:1992fk, Georges:1999, Sachdev:2010um, Sachdev:2010uj}, has emerged as a useful laboratory to understand the origins of AdS/CFT duality.
Related models have been studied \cite{Danshita:2016xbo, Erdmenger:2015xpq} with extensions \cite{Gross:2016kjj, Gu:2016oyy, Berkooz:2016cvq, Fu:2016vas, Fu:2016yrv, Garcia-Alvarez:2016wem, Hartnoll:2016mdv, Nishinaka:2016nxg, Turiaci:2017zwd, Jian:2017unn, Chew:2017xuo}
and generalizations in the form of tensor type models \cite{Witten:2016iux, Gurau:2016lzk, Klebanov:2016xxf, Peng:2016mxj, Ferrari:2017ryl, Itoyama:2017emp, Peng:2017kro} .
Interesting random matrix theory interpretations have been studied in \cite{You:2016ldz, Garcia-Garcia:2016mno, Cotler:2016fpe, Liu:2016rdi, Krishnan:2016bvg, Garcia-Garcia:2017pzl, Li:2017hdt}.
The model is notable for several reasons. It features an emergent  reparametrization and conformal invariance in the IR. The out-of-time-order correlators exhibit quantum chaos, with a Lypunov exponent characteristic of black holes, thus providing an example of the butterfly effect
\cite{Shenker:2013pqa, Leichenauer:2014nxa, Shenker:2014cwa, Maldacena:2015waa, Polchinski:2015cea, Caputa:2016tgt, Gu:2016hoy, Perlmutter:2016pkf, Anninos:2016szt, Turiaci:2016cvo}.

Like vector models, the SYK model is solvable at large $N$. 
Vector models, in general, at large $N$ can be expressed in terms of bi-local fields, and it was proposed in \cite{Das:2003vw} that these bi-local fields in fact provide a bulk construction of the dual  higher spin theory \cite{Klebanov:2002ja}, with the pair of coordinates in the bi-local combining to provide the coordinates of the emergent AdS space-time. 

The simplest proposal of \cite{Das:2003vw} was implemented nontrivially  in three dimensions giving an understanding of bulk higher spin fields \cite{Koch:2010cy, Koch:2014mxa, Koch:2014aqa}.
In the one dimensional SYK case \cite{Jevicki:2016bwu,Jevicki:2016ito} such bulk mapping is realized in its simplest form, with the bi-local times mapped to AdS$_2$ space-time, thus providing an elementary example (in addition to the $c=1$ matrix model \cite{Das:1990kaa}) of how a Large $N$ quantum mechanical model grows an additional dimension.

Nevertheless, and despite great interest, the precise bulk dual of the SYK model is still ununderstood. It has been conjectured in \cite{Jensen:2016pah, Maldacena:2016upp, Engelsoy:2016xyb,Forste:2017kwy} that the gravity sector of this model is the Jackiw-Teitelboim model \cite{Teitelboim:1983ux, jackiw} of dilaton-gravity with a negative cosmological constant, studied in \cite{Almheiri:2014cka}, while \cite{Mandal:2017thl} provides strong evidence that it is actually Liouville theory.
(See also \cite{Cvetic:2016eiv, Diaz:2016kkn, Hashimoto:2016dfz, Blake:2016jnn, Mezei:2017kmw, Diaz:2017cnz})
It is also known that the matter sector contains an infinite tower of particles \cite{Polchinski:2016xgd, Maldacena:2016hyu, Jevicki:2016bwu}.
Recently, couplings of these particles have been computed by calculating six point functions in the SYK model \cite{Gross:2017hcz}.

In this paper, we provide a three dimensional interpretation of the
bulk theory. The zero temperature SYK model
corresponds to a background AdS$_2 \times I$, where $I = S^1/Z_2$ is a finite
interval whose size needs to be suitably chosen.
There is a single scalar field coupled to gravity, whose mass is equal to the Breitenlohner-Freedman bound \cite{Breitenlohner:1982bm} of
AdS$_2$. The scalar field satisfies Dirichlet boundary conditions at the ends and feels an external delta function potential at the middle of the interval. 
Alternatively, one can consider half of the interval with Dirichlet condition at one end, and a nontrivial boundary condition determining the derivative of the field at the other end.
\footnote{We thank Edward Witten for a clarification on this point.}
The background can be thought of as coming from the
near-horizon geometry of an extremal charged black hole which
reduces the gravity sector to Jackiw-Teitelboim model with the metric
in the third direction becoming the dilaton of the latter model
\cite{Maldacena:2016upp}. The strong coupling limit of the SYK model
corresponds to a trivial metric in the third direction, while at
finite coupling this acquires a dependence on the AdS$_2$ spatial coordinate.
With a suitable choice of the size of the interval $L$ and the strength of the delta function potential $V$
we show that at strong coupling,  
(i) the spectrum of the Kaluza-Klein (KK) modes of the scalar is precisely the
spectrum of the SYK model and (ii) the two point function
\footnote{Note that this two point function is not the same as the standard AdS$_2$ propagator. We thank Juan Maldacena for discussions about this point.} with both
points at the center of the interval is in precise agreement with the
strong coupling bi-local propagator, using the simplest identification
of the AdS coordinates proposed in \cite{Das:2003vw}. For finite
coupling, we adopt the proposal of \cite{Maldacena:2016upp,
  Engelsoy:2016xyb}, and show that to order $1/J$, the poles of the propagator shift in a manner consistent with the explicit results in \cite{Maldacena:2016hyu}. 

In section \ref{sec:overview}, we review relevant aspects of the bilocal formulation of the model.
In section \ref{sec:3dinterpret}, we discuss the three dimensional interpretation.
Section \ref{sec:conclusions} contains some concluding remarks.

\section{Overview of SYK}
\label{sec:overview}
In this section, we will give a brief review of the Large $N$ formalism and results along \cite{Jevicki:2016bwu, Jevicki:2016ito}.
The Sachdev-Ye-Kitaev model \cite{Kitaev:2015} is a quantum mechanical many body system with all-to-all interactions on fermionic $N$ sites ($N \gg 1$), represented by the Hamiltonian
	\begin{equation}
		H \, = \, \frac{1}{4!} \sum_{i,j,k,l=1}^N J_{ijkl} \, \chi_i \, \chi_j \, \chi_k \, \chi_l \, ,
	\label{Hamiltonian}
	\end{equation}
where $\chi_i$ are Majorana fermions, which satisfy $\{ \chi_i, \chi_j \} = \delta_{ij}$.
The coupling constant $J_{ijkl}$ are random with a Gaussian
distribution with width $J$. The generalization to analogous $q$-point interacting model is straightforward \cite{Kitaev:2015,Maldacena:2016hyu}.
After the disorder averaging for the random coupling $J_{ijkl}$, there
is only one effective coupling $J$ in the effective action. The model
is usually treated by replica method. One does not expect a spin glass
state in this model \cite{Sachdev:2015efa} so that we can restrict to
the replica diagonal subspace \cite{Jevicki:2016bwu}.
The Large $N$ theory is simply represented through a (replica diagonal) bi-local collective field:
	\begin{equation}
		\Psi(t_1, t_2) \, \equiv \, \frac{1}{N} \sum_{i=1}^N \chi_i(t_1) \chi_i(t_2) \, ,
	\end{equation}
where we have suppressed the replica index. The corresponding path-integral is
	\begin{equation}
		Z \, = \, \int \prod_{t_1, t_2} \mathcal{D}\Psi(t_1, t_2) \ \mu(\Psi) \, e^{-S_{\rm col}[\Psi]} \, , 
	\label{eq:collective partition function}
	\end{equation}
where $S_{\rm col}$ is the collective action:
	\begin{equation}
		S_{\rm col}[\Psi] \, = \, \frac{N}{2} \int dt \, \Big[ \partial_t \Psi(t, t')\Big]_{t' = t} \, + \, \frac{N}{2} \, {\rm Tr} \log \Psi \, - \, \frac{J^2N}{2q} \int dt_1 dt_2 \, \Psi^q(t_1, t_2) \, .
	\label{S_col}
	\end{equation}
Here the trace term comes from a Jacobian factor due to the change of path-integral variable, and the trace is taken over the bi-local time.
One also has an appropriate order $\mathcal{O}(N^0)$ measure $\mu$.
This action being of order $N$ gives a systematic $G=1/N$ expansion, while the measure $\mu$ found as in \cite{Jevicki:2014mfa} begins to contribute at one-loop level (in $1/N$).
Other formulations can be employed using two bi-local fields. These can be seen to reduce to $S_{\rm col}$ after elimination.

In the above action, the first linear term represents a conformal breaking term,
while the other terms respect conformal symmetry. In the IR limit with strong coupling $|t|J \gg 1$, the collective action is reduced to the critical action 
	\begin{equation}
		S_{\rm c}[\Psi] \, = \, \frac{N}{2} \, {\rm Tr} \log \Psi \, - \, \frac{J^2N}{2q} \int dt_1 dt_2 \, \Psi^q(t_1, t_2) \, ,
	\label{S_c}
	\end{equation}
which exhibits an emergent  reparametrization symmetry
	\begin{equation}
		\Psi(t_1, t_2) \, \to \, \Psi_f(t_1, t_2) \, = \, \Big| f'(t_1) f'(t_2) \Big|^{\frac{1}{q}} \, \Psi(f(t_1), f(t_2)) \, ,
	\label{reparametrization}
	\end{equation}
with an arbitrary function $f(t)$.
This symmetry is responsible for the appearance of zero modes in the strict IR critical theory.
This problem was addressed in \cite{Jevicki:2016bwu} with analog of the quantization of extended systems with symmetry modes \cite{Gervais:1975pa}.
The above symmetry mode representing time reparametrization can be elevated to a dynamical variable introduced according to \cite{Gervais:1975yg}
through the Faddeev-Popov method, leading to a Schwarzian action 
for this variable  \cite{Jevicki:2016ito}  proposed by Kitaev, and established first at quadratic level in \cite{Maldacena:2016hyu}:
	\begin{equation}
		S[f] \, = \, - \, \frac{N\alpha}{24\pi J} \int dt \, \left[ \, \frac{f'''(t)}{f'(t)} \, - \, \frac{3}{2} \, \left( \frac{f''(t)}{f'(t)} \right)^2 \, \right] \, ,
	\label{S[f]}
	\end{equation}
where the coefficient $\alpha \, = \, - 12 \pi B_1 \gamma$, with $B_1$ representing the strength of the first order correction,
established in numerical studies by Maldacena and Stanford \cite{Maldacena:2016hyu}.
The details of the non-linear evaluation are give in \cite{Jevicki:2016ito}.

For the rest of this paper we proceed with $q=4$.
Fluctuations around the critical IR background can be studied by
expanding the bi-local field as \cite{Jevicki:2016bwu}
	\begin{equation}
		\Psi(t_1, t_2) \, = \, \Psi_0(t_1, t_2) \, + \, \sqrt{\frac{2}{N}} \ \eta(t_1, t_2) \, ,
	\end{equation}
where $\eta$ is the fluctuation and the critical IR background solution is given by
	\begin{equation}
		\Psi_0(t_1, t_2) \, = \, \left( \frac{1}{4\pi J^2} \right)^{\frac{1}{4}} \, \frac{{\rm sgn}(t_{12})}{\sqrt{|t_{12}|}} \, ,
	\end{equation}
where we defined $t_{ij}\equiv t_i -t_j$.
With a simple coordinate transformation
	\begin{equation}
		t \, = \, \frac{1}{2} \, (t_1 + t_2) \, , \qquad z \,
                = \, \frac{1}{2} \, (t_1 - t_2) \, ,
	\label{1-1}
	\end{equation}
the bi-local field $\eta(t_1, t_2)$ 
	\begin{equation}
		\eta(t_1, t_2) \, \equiv  \, \Phi(t, z) \, ,
	\end{equation}
can be then considered as a field in two dimensions $(t,z)$.
Expand the fluctuation field as
	\begin{equation}
		\Phi(t,z) = \sum_{\nu,\omega} \tilde{\Phi}_{\nu, \omega} u_{\nu, \omega}(t, z)
	\end{equation}
in a complete orthonormal basis
	\begin{equation}
		u_{\nu, \omega}(t, z) \, = \, {\rm sgn}(z) \, e^{i \omega t} \, Z_{\nu}(|\omega z|) \, ,
	\end{equation}
with $Z_{\nu}$ are a complete set of modes given in Eq.(\ref{Z_nu}), which diagonalizes the quadratic kernel \cite{Polchinski:2016xgd}.
Then, the quadratic action can be written as 
	\begin{equation}
		S_{(2)} \, = \, \frac{3J}{32\sqrt{\pi}} \sum_{\nu, \omega} \, N_{\nu} \, \tilde{\Phi}_{\nu, \omega} \big( \tilde{g}(\nu) - 1 \big) \tilde{\Phi}_{\nu, \omega} \, ,
	\label{eq:quadratic action}
	\end{equation}
where the normalization factor $N_{\nu}$ is 
\begin{align}
		N_{\nu} \, = \,
		\begin{cases}
			(2\nu)^{-1}  &{\rm for}\ \nu=3/2+2n \\
			2\nu^{-1}\sin\pi\nu \quad &{\rm for}\ \nu=ir \, ,
		\end{cases}
	\label{N_nu}
	\end{align}
and the kernel is given by
	\begin{equation}
		\tilde{g}(\nu) \, = \, - \, \frac{2\nu}{3} \, \cot\left( \frac{\pi \nu}{2} \right) \, .
	\label{tildeg(nu)}
	\end{equation}
After a field redefinition  \cite{Jevicki:2016bwu} the effective action can be written as
	\begin{equation}
		S^{\text{eff}}_m \, = \, \frac{1}{2}  \frac{3J}{8\sqrt{\pi}} \int dt \int_0^\infty {dz\over z} \, \Phi(t,z) \left[ \widetilde{g}(\sqrt{\mathsf{D}_{\rm B}} )-1 \right] \Phi(t,z) \, ,
	\end{equation}
featuring the Bessel operator 
	\begin{equation}
		\mathsf{D}_{\rm B} \, \equiv \, z^2\partial_z^2+z\partial_z -z^2\partial_t^2 \, .
	\end{equation}
The operator $\mathsf{D}_{\rm B}$ is in fact closely related to 
the laplacian on AdS$_2$,
\begin{equation}
\nabla_{AdS_2} \Phi = \sqrt{z} \mathsf{D}_{\rm B}(\frac{1}{\sqrt{z}}
\Phi) -\frac{1}{4} \Phi
\end{equation}
where $t$ and $z$ in (\ref{1-1}) are the Poincare coordinates in
AdS$_2$. This, therefore,  realizes the naive form of the proposal of
\cite{Das:2003vw}. However the action for $\Phi(t,z)$ is
non-polynomial in derivatives.

To understand the implications of this, consider the bi-local propagator, first evaluated in \cite{Polchinski:2016xgd}.
From the above effective action, one has that the poles  are determined  as solutions of  $\tilde{g}(\nu)=1$, they represent a sequence denoted by $p_m$ as
	\begin{equation}
		\frac{2p_m}{3} \, = \, - \, \tan\left( \frac{\pi p_m}{2} \right) \, , \qquad 2m+1 < p_m <2m+2 \qquad (m=0, 1, 2, \cdots )
	\label{ploes p_m}
	\end{equation}
Therefore, the bi-local propagator  is written as residues of $\nu=p_m$ poles as
	\begin{equation}
		\mathcal{D}(t, z; t', z') \, = \, - \, \frac{32 \pi^{\frac{3}{2}}}{3J} \int_{-\infty}^{\infty} d\omega \, e^{-i \omega(t-t')} \sum_{m=1}^{\infty} \, R(p_m) \
		\frac{Z_{-p_m}(|\omega| z^>) J_{p_m}(|\omega| z^<)}{N_{p_m}} \, ,
	\label{bi-local propagator}
	\end{equation}
where $z^{>}(z^{<})$ is the greater (smaller) number among $z$ and $z'$. The residue function is defined by
	\begin{equation}
		R(p_m) \, \equiv \, {\rm Res} \left( \frac{1}{\tilde{g}(\nu) - 1} \right) \bigg|_{\nu=p_m} = \, \frac{3p_m^2}{[p_m^2 + (3/2)^2][\pi p_m - \sin(\pi p_m)]} \, .
	\label{R(p_m)}
	\end{equation}
Since that $p_m$ are zeros of $\widetilde{g}(\nu)-1$, near each pole $p_m$, we can approximate as
	\begin{equation}
		\widetilde{g}(\nu)-1 \, \approx \, \left[\nu^2-(p_m)^2\right]f_m \, ,
	\end{equation}
where $f_m $ can be determined from residue of $1/(\widetilde{g}(\nu)-1)$ at $\nu = p_m$.
Explicitly evaluating these residues, the inverse kernel is written as an exact expansion
 	\begin{equation}
		\frac{1}{\tilde{g}(\nu) - 1} \, = \, \sum_{m=1}^{\infty} \, \frac{6 \, p_m^3}{[p_m^2 + (3/2)^2][\pi p_m - \sin(\pi p_m)]} \, \left( \frac{1}{\nu^2 - p_m^2} \right) \, .
	\label{Laurent expansion}
\end{equation}
The effective action near a pole labelled by $m$ is that of a scalar field with mass, $M^2_m=p_m^2-{1\over 4}$, ($m>0$) in AdS$_2$:
	\begin{equation}
		S^{\rm eff}_m \, = \, \frac{1}{2} \int \sqrt{-g} \, d^2x \left[ -g^{\mu\nu} \partial_\mu\phi_m \partial_\nu\phi_m -\left(p_m^2-{1\over 4}\right)\phi_m^2 \right] \, ,
	\label{eq:effective action for scalar field 1}
	\end{equation}
where the metric $g_{\mu\nu}$ is given by $g_{\mu\nu}=\mbox{diag}(-1/z^2,1/z^2)$.
It is clear from the above analysis that a spectrum of a sequence of 2D scalars, with growing conformal dimensions is being packed into a single bi-local field.
In other words the bi-local representation effectively packs an infinite product of AdS Laplacians with growing masses.
An illustration of how this can happen is given in the appendix \ref{sec:nonpolynomial}, relating to the scheme of Ostrogradsky.
It is this feature which leads to the suggestion that the theory should be represented by an enlarged number of fields, or equivalently by an extra Kaluza-Klein dimension.

For finite coupling, the poles of the propagator is shifted. This has been calculated by \cite{Maldacena:2016hyu} in a $1/J$ expansion. 

\section{3D Interpretation}
\label{sec:3dinterpret}

According to \cite{Maldacena:2016upp} and \cite{Engelsoy:2016xyb}, the
bulk dual of the SYK model involves Jackiw-Teitelboim theory of two
dimensional dilaton gravity, whose action is given by (upto usual
boundary terms)
\ben
S_{JT} = -\frac{1}{16\pi G} \int \sqrt{-g} \Big[ \phi(R + 2) - 2\phi_0 \Big]  \, ,
\een
where $\phi_0$ is a constant, and $\phi$ is a dilaton field. The zero
temperature background is given by AdS$_2$ with a metric
\ben
ds^2 = \frac{1}{z^2}[-dt^2 + dz^2]
\een
and a dilaton  
\ben
\phi(z) = \phi_0 + \frac{a}{z}
\label{dilaton}
\een
where $a$ is a parameter which scales as $1/J$. In the following we will choose, without
loss of generality, $\phi_0 = 1$.

This action can be thought as arising from a higher dimensional system
which has extremal black holes, and the AdS$_2$ is the near horizon
geometry \cite{Maldacena:2016upp}. The three dimensional metric, with the dilaton being the third direction, is given by
\ben
ds^2 = \frac{1}{z^2}\big[ -dt^2 + dz^2 \big] \, + \, \left(1+\frac{a}{z}\right)^2 dy^2 \, .
\label{1-2}
\een
This is in fact the near-horizon geometry of a charged extremal BTZ
black hole.

\subsection{Kaluza-Klein Decomposition}
\label{sec:Kaluza-Klein Decomposition}

We will now show that the infinite sequence of poles in the previous
section from the Kaluza-Klein tower of a single scalar in a three dimensional metric
(\ref{1-2}) where the direction $y$ is an interval $ -L < y < L$. The
action of the scalar is
	\begin{equation}
		S \, = \, \frac{1}{2} \int d^3x \sqrt{-g} \Big[ - g^{\mu\nu} \, \partial_{\mu} \Phi \, \partial_{\nu} \Phi \, - \, m_0^2 \, \Phi^2 \, - \, V(y) \Phi^2 \Big] \, ,
	\label{action2}
	\end{equation}
where $V(y) = V \delta(y)$, with the constant $V$ and the size $L$ to be determined.
This is similar to Horava-Witten compactification on $S^1/Z_2$ \cite{Horava:1995qa} with an additional delta function potential.
\footnote{See also \cite{Georgi:2000ks,Carena:2002me}. We are grateful to Cheng Peng for bringing this to our attention.}
The scalar satisfies Dirichlet boundary conditions at the ends of the interval.

We now proceed to decompose the 3D theory into 2 dimensional modes. Using Fourier transform for the $t$ coordinate:
	\begin{equation}
		\Phi(t, z, y) \, = \, \int \frac{d\omega}{2\pi} \, e^{-i\omega t} \, \chi_{\omega}(z, y) \, ,
	\end{equation}
one can rewrite the action (\ref{action2}) in the form of
	\begin{equation}
		S \, = \, \frac{1}{2} \int dz dy \int \frac{d\omega}{2\pi} \ \chi_{-\omega} \, (\mathcal{D}_0 + \mathcal{D}_1) \, \chi_{\omega}\, ,
	\end{equation}
where $\mathcal{D}_0$ is the $a$-independent part and $\mathcal{D}_1$ is linear in $a$:
	\begin{align}
		\mathcal{D}_0 \, &= \, \partial_z^2 \, + \, \omega^2 \, - \, \frac{m_0^2}{z^2} \, + \, \frac{1}{z^2} \Big( \partial_y^2 - V(y) \Big) \, , \nonumber\\
		\mathcal{D}_1 \, &= \, \frac{a}{z} \left[ \, \partial_z^2 \, - \, \frac{1}{z} \, \partial_z \, + \, \omega^2 \, - \, \frac{m_0^2}{z^2} \, - \, \frac{1}{z^2} \Big( \partial_y^2 + V(y) \Big) \right] \, .
	\end{align}
Here, we neglected higher order contributions of $a$.
The eigenfunctions of $\mathcal{D}_0$ can be clearly written in the form
	\begin{equation}
		\chi_{\omega}(z, y) \, = \, \chi_{\omega}(z) \, f_k(y) \, .
	\end{equation}
Then $f_k(y)$ is an eigenfunction of the ${\rm Schr\ddot{o}dinger}$ operator $-\partial_y^2 + V(y)$ with eigenvalue $k^2$.
This is a well known Schrodinger problem: the eigenfunctions and the eigenvalues are presented in detail in Appendix \ref{app:schrodinger equation}.

After solving this part, the kernels are reduced to 
	\begin{equation}
		\mathcal{D}_0 \, = \, \partial_z^2 \, + \, \omega^2 \, - \, \left( \frac{m_0^2+p_m^2}{z^2} \right) \, , \qquad
		\mathcal{D}_1 \, = \, \frac{a}{z} \left[ \, \partial_z^2 \, - \, \frac{1}{z} \, \partial_z \, + \, \omega^2 \, - \, \left( \frac{m_0^2-q_m^2}{z^2} \right) \right] \, ,
	\label{D0&D1}
	\end{equation}
where $p_m$ are the solutions of 
\ben
-(2/V)k=\tan(kL)
\label{3-1}
\een
while $q_m$ are the expectation values of $-\partial_y^2 - V(y)$ operator respect to $f_{p_m}$.
If we choose $V=3$ and $L=\frac{\pi}{2}$ the solutions of (\ref{3-1}) agree precisely with the strong coupling spectrum of the SYK model given by ${\tilde{g}}(\nu) = 1$, as is clear from (\ref{eq:quadratic action}) and (\ref{tildeg(nu)}). This is our main observation.

For these values of $V$, $L$, the propagator $G$ is determined by the Green's equation of $\mathcal{D}$.
We now use the perturbation theory to evaluate it. This will then be compared with the corresponding propagator of the bi-local SYK theory.

\subsection{Evaluation of $G^{(0)}$}
\label{sec:evaluation of G^0}
We start by determining the leading, zero-th order $G^{(0)}$ propagator obeying
	\begin{equation}
		\mathcal{D}_0 \, G^{(0)}_{\omega, \omega'}(z,y; z',y') \, = \, -  \, \delta(z-z')\delta (y-y')\delta(\omega + \omega') \, .
	\label{G^0 Greens eq}
	\end{equation}
We first separate the scaling part of the propagator by $G^{(0)}=\sqrt{z}\, \widetilde{G}^{(0)}$ and multiplying $z^2$. Expanding in a basis of eigenfunctions $f_k(y)$,
\ben
\widetilde{G}^{(0)}(z,y,\omega;z',y',\omega') = \sum_{k,k'} f_k(y) f_{k'}(y') \widetilde{G}^{(0)}_{\omega,k;\omega',k'} (z;z')
\een
The Green's function $\widetilde{G}^{(0)}_{\omega,k;\omega',k'} (z,z')$ is clearly proportional to $\delta (k-k')$ and satisfies the equation
	\begin{equation}
		\Big[ \, z^2 \, \partial_z^2 \, + \, z \, \partial_z \, + \, \omega^2 \, z^2 \, - \, \nu_0^2 \, \Big] \, \widetilde{G}^{(0)}_{\omega, k; \omega', k'}(z; z')
		\, = \, - \, z^{\frac{3}{2}} \, \delta(z-z') \delta(\omega + \omega') \delta (k-k') \, .
\label{3-12}
	\end{equation}
where we have defined 
\ben
\nu_0^2\equiv k^2+m_0^2+1/4.
\een
The operator which appears in  (\ref{3-12}) is the Bessel operator. Thus the Green's function can be expanded in the complete orthonormal basis.
For this, we use the same basis form $Z_{\nu}$ as in the SYK evaluation
\footnote{This represents a modified set of wavefunctions with boundary conditions at $z \to \infty$ in contrast to the standard AdS wavefunctions.}:
	\begin{equation}
		\widetilde{G}^{(0)}_{\omega, k; -\omega, k}(z; z') \, = \, \int d\nu \, \widetilde{g}_{\nu}^{(0)}(z') \, Z_{\nu}(|\omega z|) \, .
	\end{equation}
Then, substituting this expansion into the Green's equation (\ref{G^0 Greens eq}) and using Eqs.(\ref{completeness-Z_nu}) and (\ref{Bessel eq}),
one can fix the coefficient $\widetilde{g}_{\nu}^{(0)}$.
Finally, the $\nu$-integral form of the propagator is given by
	\begin{equation}
		G^{(0)}_{\omega, k; -\omega, k}(z; z') \, = \, - \, |zz'|^{\frac{1}{2}} \int \frac{d\nu}{N_{\nu}} \, \frac{Z^*_{\nu}(|\omega z|) \, Z_{\nu}(|\omega z'|)}{\nu^2 - \nu_0^2} \, .
	\end{equation}
We now note that if we choose $m_0^2 = -1/4$, which is the BF bound of AdS$_2$, we have $\nu_0^2 = p_m^2$, and the equation which determine $p_m$, (\ref{3-1}) is precisely the equation which determines the spectrum of the SYK theory found in \cite{Polchinski:2016xgd, Jevicki:2016bwu}. With this choice, the real space zeroth order propagator in three dimensions is 
	\begin{equation}
		G^{(0)}(t,z,y; t',z',y') \, = \, - \, |zz'|^{\frac{1}{2}} \sum_{m=0}^{\infty} f_{p_m}(y) f_{p_m}(y') \int \frac{d\omega}{2\pi} \, e^{-i\omega(t-t')}
		\int \frac{d\nu}{N_{\nu}} \, \frac{Z^*_{\nu}(|\omega z|) \, Z_{\nu}(|\omega z'|)}{\nu^2 - p_m^2} \, .
	\label{G^0-nu}
	\end{equation}

We now show that the above propagator with $y = y'=0$ is in exact agreement with the bi-local propagator of the SYK model. The Green's function with these end points is
	\begin{equation}
		G^{(0)}(t,z,0; t',z',0) \, = \, - \, |zz'|^{\frac{1}{2}} \sum_{m=0}^{\infty} \, C(p_m) \int \frac{d\omega}{2\pi} \, e^{-i\omega(t-t')} \,
		\int \frac{d\nu}{N_{\nu}} \, \frac{Z^*_{\nu}(|\omega z|) \, Z_{\nu}(|\omega z'|)}{\nu^2 - p_m^2} \, ,
	\label{G^0 at y=0}
	\end{equation}
where we have defined
	\begin{equation}
		C(p_m) \, \equiv \, f_{p_m}(0) f_{p_m}(0) \, = \, B_m^2 \, \frac{p_m^2}{p_m^2+(3/2)^2}  \, = \, \frac{2p_m^3}{[p_m^2 + (3/2)^2][\pi p_m - \sin(\pi p_m)]} \, .
	\end{equation}
Now we note that Kaluza-Klein wave function coefficient coincides in detail with the SYK one, namely:
	\begin{equation}
		C(p_m) \, = \, \frac{2p_m}{3} \, R(p_m) \, , 
	\end{equation}
where $R(p_m)$ was given  in Eq.(\ref{R(p_m)}).

As in Eq.(\ref{completeness-Z_nu}),
the integration of $\nu$ is a short-hand notation which denotes a summation of $\nu=3/2+2n$, ($n=0, 1, 2 \cdots$) and an integral of $\nu=ir$, ($0<r<\infty$).
The sum over these discrete values of $\nu$ and the integral over the continuous values can be now performed exactly as in the calculation of the SYK bi-local propagator \cite{Jevicki:2016bwu}.
Closing the contour for the continuous integral in Re($\nu$)$\to \infty$, one finds that there are two types of poles inside of this contour.
(1): $\nu=2n+3/2$, ($n=0, 1, 2, \cdots$), and (2): $\nu=p_m$, ($m=0, 1, 2, \cdots$).
The contributions of the former type of poles precisely cancel with the contribution from the discrete sum over $n$.
Details of the evaluation which explicitly shows the cancelation are presented in Appendix \ref{app:evaluation of the contour integral}.
Therefore, the final remaining contribution is just written as residues of $\nu=p_m$ poles as
	\begin{equation}
		G^{(0)}(t,z,0; t',z',0) \, = \, \frac{1}{3} \, |zz'|^{\frac{1}{2}} \sum_{m=0}^{\infty} \int_{-\infty}^{\infty} d\omega \, e^{-i\omega(t-t')} \, R(p_m) \
		\frac{Z_{-p_m}(|\omega| z^>) J_{p_m}(|\omega| z^<)}{N_{p_m}} \, .
	\label{G^0 at y=0-result}
	\end{equation}
Altogether we have shown that $y=0$ mode 3D propagator is in precise agreement with the $q=4$ SYK bi-local propagator at large $J$ given in Eq.(\ref{bi-local propagator}).
The propagator is a sum of non-standard propagators in AdS$_2$.
While it vanishes on the boundary, the boundary conditions at the horizon are different from that of the standard propagator in AdS.

\subsection{First Order Eigenvalue Shift}
\label{sec:first order eigenvalue shift}
In this section, we study the first order eigenvalue shift due to $\mathcal{D}_1$ by treating this operator as a perturbation onto the $\mathcal{D}_0$ operator.
The result will confirm the duality $a = 1/J$, where $a$ is defined in the dilaton background (\ref{dilaton}) and $J$ is the coupling constant in the SYK model.

Since the $t$ and $y$ directions are trivial, let us start with the kernels already solved for these two directions given in Eq.(\ref{D0&D1}).
The eigenfunction of $\mathcal{D}_0$ operator is
	\begin{equation}
		|z|^{\frac{1}{2}} \, Z_{\nu}(|\omega z|) \, ,
	\label{D0-eigenfunction}
	\end{equation}
and using the orthogonality condition (\ref{orthogonality}), its matrix element in the $\nu$ space is found as
	\begin{equation}
		N_{\nu} \Big[ \nu^2 - (m_0^2 + p_m^2 + \tfrac{1}{4}) \Big] \, \delta_{\nu, \nu'} \, .
	\end{equation}

Now following the first order perturbation theory, we are going to determine the first order eigenvalue shift.
Using the Bessel equation, the action of $\mathcal{D}_1$ on the $\mathcal{D}_0$ eigenfunction (\ref{D0-eigenfunction}) is found as
	\begin{equation}
		\mathcal{D}_1 \, |z|^{\frac{1}{2}} \, Z_{\nu}(|\omega z|)
		\, = \, \frac{a}{|z|^{\frac{1}{2}}} \left[ \, \frac{\partial_z}{z} \, - \, \left( \frac{m_0^2-q_m^2+\tfrac{3}{4}}{z^2} \right) \right] \, Z_{\nu}(|\omega z|) \, .
	\end{equation}
For the derivative term, we use the Bessel function identity (for example, see 8.472 of \cite{Gradshteyn:1994})
	\begin{equation}
		\partial_x J_{\nu}(x) \, = \, \pm \, J_{\nu \mp 1}(x) \, \mp \, \frac{\nu}{x} \, J_{\nu}(x) \, ,
	\end{equation}
to obtain
	\begin{equation}
		\partial_z \, Z_{\nu}(|\omega z|) \, = \, \frac{\nu}{|z|} \, Z_{\nu}(|\omega z|) \, - \, |\omega| \Big[ J_{\nu+1}(|\omega z|) \, - \, \xi_{\nu} \, J_{-\nu-1}(|\omega z|) \Big] \, .
	\end{equation}
Therefore, now the matrix element is determined by integrals
	\begin{align}
		\int_0^{\infty} dz \, |z|^{\frac{1}{2}} \, Z_{\nu'}^*(|\omega z|) \mathcal{D}_1 \, |z|^{\frac{1}{2}} \, Z_{\nu}(|\omega z|)
		\, &= \, a \Big[ \nu - \left( m_0^2-q_m^2+\tfrac{3}{4}\right) \Big] \int_0^{\infty} dz \, \frac{Z_{\nu'}^*(|\omega z|)Z_{\nu}(|\omega z|)}{z^2} \nonumber\\
		&\quad - \, a |\omega| \int_0^{\infty} dz \, \frac{Z_{\nu'}^*(|\omega z|)}{z} \, \Big[ J_{\nu+1}(|\omega z|) \, - \, \xi_{\nu} \, J_{-\nu-1}(|\omega z|) \Big] \, .
	\end{align}
For the continuous mode ($\nu=ir$), the integrals might be hard to evaluate.
In the following, we restrict ourself to the real discrete mode $\nu=3/2+2n$.
In such case, $\xi_{\nu}=0$. Therefore, the linear combination of the Bessel function is reduced to a single Bessel function as $Z_{\nu}(x) = J_{\nu}(x)$.
Since 
	\begin{align}
		\int_0^{\infty} dx \, \frac{J_{\alpha}(x) J_{\beta}(x)}{x} \, &= \, \frac{2}{\pi} \, \frac{\sin\big[ \tfrac{\pi}{2}(\alpha- \beta) \big]}{\alpha^2 - \beta^2} \, , \hspace{91pt}
		[{\rm Re}(\alpha), {\rm Re}(\beta) > 0] \nonumber\\
		\int_0^{\infty} dx \, \frac{J_{\alpha}(x)J_{\beta}(x)}{x^2}
		\, &= \, \frac{4}{\pi} \, \frac{\sin\big[ \tfrac{\pi}{2}(\alpha- \beta-1) \big]}{\big[ (\alpha + \beta)^2 -1 \big]\big[ (\alpha - \beta)^2 -1 \big]} \, , \qquad [{\rm Re}(\alpha), {\rm Re}(\beta) > 1]
	\end{align}
we have now found the matrix element for the discrete mode is given by
	\begin{equation}
		\frac{2a |\omega|}{\pi} \frac{\sin\big[ \tfrac{\pi}{2}(\nu- \nu'-1) \big]}{(\nu+1)^2 - \nu'^2} \left[ \frac{2\big[ \nu - (m_0^2 - q_m^2+\tfrac{3}{4})\big]}{(\nu-1)^2 - \nu'^2} \, - \, 1 \right] \, .
	\end{equation}

Next, let us focus on the zero mode ($\nu=\nu'=3/2$) eigenvalue.
In the above formula, taking the bare mass to the BF bound: $m_0^2 = -1/4$ as before, the zero mode first order eigenvalue shift is found as
	\begin{equation}
		\frac{a |\omega|}{2\pi} \, (2 + q_0^2) \, .
	\label{zero mode eigenvalue shift}
	\end{equation}
Now, we compare this result with the $1/J$ first order eigenvalue shift of the SYK model, which is for the zero mode found in \cite{Maldacena:2016hyu} as
	\begin{equation}
		k(2, \omega) \, = \, 1 \, - \, \frac{\alpha_K|\omega|}{2 \pi \mathcal{J}} \, + \, \cdots \, , \qquad ({\rm zero~ temperature})
	\end{equation}
where $\alpha_K\approx 2.852$ for $q=4$.
The $\omega$-dependence of our result (\ref{zero mode eigenvalue shift}) thus agrees with that of the SYK model.
Furthermore, this comparison confirms the duality $a = 1/J$.

Finally, we can now complete our comparison by showing agreement for the $m=0$ mode contribution to the propagator. 
We include the first $\mathcal{O}(a)$ order shift for the pole as
	\begin{equation}
		\nu \, = \, \frac{3}{2} \, + \, \frac{a |\omega|}{6\pi} \big( 2+q_0^2 \big) \, + \, \mathcal{O}(a^2) \, .
	\end{equation}
For the zero mode part ($m=0$) of the on-shell propagator in Eq.(\ref{on-shell propagator}), the leading order is $\mathcal{O}(1/a)$.
This contribution comes from the coefficient factor of the Bessel function, which was responsible for the double pole at $\nu=3/2$.
For other $p_0$ setting them to $3/2$, we obtain the leading order contribution from the zero mode as
	\begin{equation}
		G^{(0)}_{\rm zero-mode}(t,z,0; t',z',0) \, = \, - \, \frac{9\pi}{4a} \, \frac{B_0^2}{(2+q_0^2)} \, |zz'|^{\frac{1}{2}} \int_{-\infty}^{\infty} \frac{d\omega}{|\omega|} \,
		e^{-i\omega(t-t')} J_{\frac{3}{2}}(|\omega z|) J_{\frac{3}{2}}(|\omega z'|) \, .
	\end{equation}
This agrees with the order $\mathcal{O}(J)$ contribution of the SYK bi-local propagator of Maldacena/Stanford \cite{Maldacena:2016hyu}.

\section{Conclusion}
\label{sec:conclusions}
In this paper we have provided a three dimensional perspective of the bulk dual of the SYK model. At strong coupling we showed that the spectrum and the propagator of the bi-local field can be exactly reproduced by that of a scalar field living in AdS$_2 \times S^1/Z_2$ with a delta function potential at the center. The metric on the interval in the third direction is the dilaton of Jackiw-Teitelboim theory, which is a constant at strong coupling. We also calculated the leading $1/J$ correction to the propagator which comes from the corresponding term in the metric in the third direction, and showed that form of the poles of the propagator are consistent with the results of the SYK model \cite{Maldacena:2016hyu}.

This three dimensional view is a good way of re-packaging the infinite tower of states of the SYK model.
Our analysis was done at the linearized level and the 3D gravity is only used to fix the background, as we did not treat them dynamically.
\footnote{We thank Juan Maldacena for a clarification regarding this point.}
Demonstrating full duality at the nonlinear level is an open problem.
In particular it would be interesting if the three point function of bi-locals \cite{Gross:2017hcz} has a related 3d interpretation.

\acknowledgments
We acknowledge useful conversations with Robert de Melo Koch, Animik Ghosh, Juan Maldacena, Cheng Peng, Al Shapere, Edward Witten and Junggi Yoon on the topics of this paper.
We also thank Wenbo Fu, Alexei Kitaev, Grigory Tarnopolsky and Jacobus Verbaarschot for relevant discussions on the SYK model. 
This work is supported by the Department of Energy under contract DE-SC0010010.
The work of SRD is partially supported by the National Science Foundation grant NSF-PHY-1521045.
AJ would like to thank the Galileo Galilei Institute for Theoretical Physics (GGI) for the hospitality and INFN for partial support during the completion of this work,
within the program ``New Developments in AdS3/CFT2 Holography''.
We also learned of possibly related work by Marika Taylor \cite{Taylor:2017}.

\appendix
\section{Actions non-polynomial in derivatives}
\label{sec:nonpolynomial}
To illustrate how an action which is non-polynomial in derivatives can arise let us start with the example of $N$ decoupled fields
	\begin{equation}
		\mathcal{L} \, = \, \sum_{n=1}^N \, \varphi_n \mathcal{D}_n \varphi_n \, .
	\end{equation}
One can then introduce fields
	\begin{equation}
		\varphi \, = \, \sum_{n=1}^N \, \varphi_n \, , \qquad {\rm and} \qquad \chi_n \qquad (n=1, \cdots, N-1) \, ,
	\end{equation}
so that the Lagrangian is rearranged to 
	\begin{equation}
		\mathcal{L} \, = \, \varphi \, \widehat{\mathcal{D}} \, \varphi \, + \, \sum_{n=1}^{N-1} \, \chi_n \widehat{\mathcal{D}}_n \varphi_n \, ,
	\end{equation}
where 
	\begin{align}
		\widehat{\mathcal{D}} \, &= \, \frac{\prod_{n=1}^N\mathcal{D}_n}{\sum_{n_1<\cdots<n_{N-1}}^N \mathcal{D}_{n_1} \cdots \mathcal{D}_{n_{N-1}}} \, , \nonumber\\
		\widehat{\mathcal{D}}_p \, &= \, \frac{\sum_{n_1<\cdots<n_p}^{p+1} \mathcal{D}_{n_1} \cdots \mathcal{D}_{n_p}}
		{\sum_{n_1<\cdots<n_{p-1}}^p \mathcal{D}_{n_1} \cdots \mathcal{D}_{n_{p-1}}} \, , \qquad (p=1, \cdots, N-1)
	\end{align}
which represents a transformation preserving the determinant:
	\begin{equation}
		\prod_{n=1}^N \, \mathcal{D}_n \, = \, \widehat{\mathcal{D}} \, + \, \prod_{n=1}^{N-1} \, \widehat{\mathcal{D}}_n \, .
	\end{equation}
Integrating $\chi$'s out, one eventually ends up with the effective Lagrangian
 	\begin{equation}
		\mathcal{L}_{\varphi} \, = \, \varphi \, \left( \frac{\prod_{n=1}^N\mathcal{D}_n}{\sum_{n_1<\cdots<n_{N-1}}^N \mathcal{D}_{n_1} \cdots \mathcal{D}_{n_{N-1}}} \right) \, \varphi \, .
	\end{equation}

Here all the poles are contained in the higher-order laplacian, as in Eq.(\ref{tildeg(nu)}).
The opposite procedure of going from this effective action with the $N$-th order Laplacian to the first one, requires introducing $N-1$ extra fields,
which would correspond to the scheme introduced by Ostrogradsky.

\section{Schrodinger Equation}
\label{app:schrodinger equation}
In this appendix, we consider the equation of $f(y)$, which is the ${\rm Schr\ddot{o}dinger}$ equation:
	\begin{equation}
		\Big[ - \partial_y^2 \, + \, V \delta(y) \Big] f(y) \, = \, E \, f(y) \, ,
	\label{Schrodinger-eq}
	\end{equation}
where $E$ is an eigenvalue of the equation.
Since we confined the field in $-L<y<L$, we have boundary conditions: $f(\pm L)=0$.
The continuation conditions at $y=0$ are $f(+0)=f(-0)$ and the other can be derived by integrating the ${\rm Schr\ddot{o}dinger}$ equation (\ref{Schrodinger-eq})
over ($-\varepsilon, \varepsilon$) and taking limit $\varepsilon \to 0$ as
	\begin{equation}
		f'(+0) \, - \, f'(-0) \, = \, V \, f(0) \, .
	\label{continuity}
	\end{equation}
Since the potential of the ${\rm Schr\ddot{o}dinger}$ equation is even function, the wave function is either odd or even function of $y$.

(i) odd: 
For odd parity case, a solution satisfying the boundary conditions at $y=\pm L$ is given by 
	\begin{align}
		f(y) \, = \,
		\begin{cases}
			A \sin(k(y-L)) \qquad (0 < y < L) \\
			A \sin(k(y+L)) \qquad (-L < y < 0)
		\end{cases}
	\end{align}
where $k^2=E$.
For odd parity solution, to satisfy the boundary condition $f(+0)=f(-0)$, we need $f(\pm0)=0$.
This implies that
	\begin{equation}
		k \, = \, \frac{\pi n}{L} \, , \qquad (n=1, 2, 3, \cdots)
	\end{equation}
Then, the continuity condition (\ref{continuity}) is automatically satisfied. 
The normalization constant is fixed as $A=1/\sqrt{L}$.

(ii) even: 
For even parity case, a solution satisfying the boundary conditions at $y=\pm L$ is given by 
	\begin{align}
		f(y) \, = \,
		\begin{cases}
			B \sin(k(y-L)) \qquad (0 < y < L) \\
			-B \sin(k(y+L)) \quad\, (-L < y < 0)
		\end{cases}
	\label{even psi}
	\end{align}
where $k^2=E$.
The evenness of the parity guarantees $f(-0)=f(+0)$.
So, we only need to impose the condition (\ref{continuity}) on this solution.
This condition gives an equation 
	\begin{align}
		- \frac{2}{V} \, k \, = \, \tan(kL) \, .
	\end{align}
Now we set $L=\pi/2$ and $V=3$, then we have $-(2/3)k=\tan(\pi k/2)$,
which is precisely the same transcendental equation determining poles of the $q=4$ SYK bi-local propagator (\ref{ploes p_m}).
We denote the solutions of $-(2/3)k=\tan(\pi k/2)$ by $p_m$, ($2m+1<p_m<2m+2$), ($m=0, 1, 2, \cdots$).
The normalization constant is fixed as
	\begin{equation}
		B \, = \, \sqrt{\frac{2k}{2kL-\sin(2kL)}} \ .
	\label{B}
	\end{equation}

Finally, let us prove the orthogonality of the parity even wave function (\ref{even psi}):
	\begin{equation}
		\int_{-L}^L dy \, f_m(y) f_{m'}(y) \, = \, \delta_{m, m'} \, .
	\label{orthogonality-psi}
	\end{equation}
Using the solution (\ref{even psi}) and evaluating the integral in the left-hand side, one obtains
	\begin{equation}
		B^2 \left[ \, \frac{\sin(L(k-k'))}{k-k'} \, - \, \frac{\sin(L(k+k'))}{k+k'} \, \right] \, .
	\label{integral result}
	\end{equation}
Now let's assume $k \ne k'$.
Then, the integral result can be rearranged to the form of 
	\begin{equation}
		\frac{B^2}{k^2 - k '^2} \, \cos(Lk) \cos(Lk') \, \Big[ \, k' \tan(Lk) \, - \, k \tan(Lk') \, \Big] \, = \, 0 \, ,
	\end{equation}
where the final equality is due to the relation $\tan(LK)=-2k/3$.
Next, we consider $k=k'$ case. In this case, due to the delta function identity, the result (\ref{integral result}) is reduced to 
	\begin{equation}
		B^2 \left[ \, L \, - \, \frac{\sin(2Lk)}{2k} \, \right] \, \delta_{k, k'} \, = \, \delta_{k,k'} \, ,
	\end{equation}
where for the equality we used Eq.(\ref{B}).
Therefore, now we have proven the orthogonality (\ref{orthogonality-psi}).

\section{Completeness Condition of $Z_{\nu}$}
\label{app:completeness condition of z}
In this appendix, we summarize some properties of the Bessel function $Z_{\nu}$,
which are used to determine the zero-th order propagator (\ref{G^0-nu}).
The linear combination of the Bessel functions is defined by \cite{Polchinski:2016xgd}
	\begin{equation}
		Z_{\nu}(x) \, = \, J_{\nu}(x) \, + \, \xi_{\nu} \, J_{-\nu}(x) \, , \qquad \xi_{\nu} \, = \, \frac{\tan(\pi \nu/2)+1}{\tan(\pi \nu/2)-1} \, ,
	\label{Z_nu}
	\end{equation}
which satisfies the Bessel equation
	\begin{equation}
		\left[ \, z^2 \, \partial_z^2 \, + \, z \, \partial_z \, + \, \omega^2 \, z^2 \, \right] \, Z_{\nu}(|\omega z|) \, = \, \nu^2 \, Z_{\nu}(|\omega z|) \, .
	\label{Bessel eq}
	\end{equation}

In \cite{Polchinski:2016xgd}, the orthogonality condition of the linear combination of the Bessel function $Z_{\nu}$ (\ref{Z_nu}) is given by
	\begin{equation}
		\int_0^{\infty} \frac{dx}{x} \, Z^*_{\nu}(x) \, Z_{\nu'}(x) \, = \, N_{\nu} \, \delta(\nu-\nu') \, ,
	\label{orthogonality}
	\end{equation}
where $N_{\nu}$ is defined in (\ref{N_nu}).


From this orthogonality condition, one can fix the normalization for the completeness condition of $Z_{\nu}$.
Namely, dividing each $Z_{\nu}$ by $\sqrt{N_{\nu}}$, finally we find the completeness condition as
	\begin{equation}
		\int \frac{d\nu}{N_{\nu}} \, Z_{\nu}^*(|x|) \, Z_{\nu}(|x'|) \, = \, x \, \delta(x-x') \, .
	\label{completeness-Z_nu}
	\end{equation}

\section{Evaluation of the Contour Integral}
\label{app:evaluation of the contour integral}
In this appendix, we give a detail evaluation of the continuous and the discrete sums appearing in Eq.(\ref{G^0 at y=0}).
As we defined before, the integral symbol $d\nu$ is a short-hand notation of a combination of
summation over $\nu=3/2+2n$, $(n=0, 1, 2, \cdots)$ and integration of $\nu=ir$, $(r>0)$.
Namely,
	\begin{equation}
		\int \frac{d\nu}{N_{\nu}} \, \frac{Z^*_{\nu}(|\omega z|) \, Z_{\nu}(|\omega z'|)}{\nu^2 - p_m^2} \, = \, I_1 \, + \, I_2 \, ,
	\end{equation}
with
	\begin{align}
		I_1 \, &\equiv \, \sum_{n=0}^{\infty} \, \frac{2\nu}{\nu^2-p_m^2} \, J_{\nu}(|\omega z|) \, J_{\nu}(|\omega z'|) \Big|_{\nu=\frac{3}{2}+2n} \, , \nonumber\\
		I_2 \, &\equiv \, - \int_0^{\infty} \frac{dr}{2\sinh(\pi r)} \frac{r}{r^2+p_m^2} \, Z^*_{ir}(|\omega z|) \, Z_{ir}(|\omega z'|) \, .
	\end{align}
Let us evaluate the continuous sum $I_2$ first.
Using the symmetry of the integrand, one can rewrite the integral as
	\begin{equation}
		I_2 \, = \, - \, \frac{i}{2} \int_{-i\infty}^{i\infty} \frac{d\nu}{\sin(\pi \nu)} \frac{\nu}{\nu^2-p_m^2} \,
		\Big[ J_{-\nu}(|\omega z|) \, + \, \xi_{-\nu} \, J_{\nu}(|\omega z|) \Big] \, J_{\nu}(|\omega z'|) \, .
	\end{equation}
We evaluate this integral by a contour integral on the complex $\nu$ plane by closing the contour in the Re($\nu$)$>0$ half of the complex plane if $z>z'$.
Inside of this contour, we have two types of the poles. (i) at $\nu=p_m$ coming from the coefficient factor. (ii) at $\nu=3/2+2n$, ($n=0, 1, 2, \cdots$) coming from $\xi_{-\nu}$, where $\xi_{-\nu}=\infty$.
After evaluating residues at these poles, one obtains
	\begin{align}
		I_2 \, = \, &- \, \frac{\pi}{2\sin(\pi p_m)} \, \Big[ J_{-p_m}(|\omega z|) \, + \, \xi_{-p_m} \, J_{p_m}(|\omega z|) \Big] \, J_{p_m}(|\omega z'|) \nonumber\\
		&- \, \sum_{n=0}^{\infty} \frac{2\nu}{\nu^2 - p_m^2} \, J_{\nu}(|\omega z|) \, J_{\nu}(|\omega z'|) \Big|_{\nu=\frac{3}{2}+2n} \, .
	\end{align}
Now, one can notice that the second term exactly cancels with the contribution from $I_1$.
One can also repeat the above discussion for $z'>z$ case.
Therefore, combining these two cases the total contribution is now
	\begin{equation}
		I_1 \, + \, I_2 \, = \, - \, \frac{\pi}{2\sin(\pi p_m)} \left[ J_{-p_m}(|\omega| z^>) + \left( \frac{p_m+\tfrac{3}{2}}{p_m-\tfrac{3}{2}} \right) J_{p_m}(|\omega| z^>) \right]
		J_{p_m}(|\omega| z^<) \, ,
	\end{equation}
where $z^{>}(z^{<})$ is the greater (smaller) number among $z$ and $z'$. Then, the propagator is reduced to 
	\begin{align}
		G^{(0)}(t,z,0; t',z',0) \, &= \, \frac{1}{4} \, |zz'|^{\frac{1}{2}} \sum_{m=0}^{\infty} \int_{-\infty}^{\infty} d\omega \, e^{-i\omega(t-t')}
		\frac{B_m^2}{\sin(\pi p_m)} \, \frac{p_m^2}{p_m^2+(3/2)^2} \nonumber\\
		&\hspace{50pt} \times \left[ J_{-p_m}(|\omega| z^>) + \left( \frac{p_m+\tfrac{3}{2}}{p_m-\tfrac{3}{2}} \right) J_{p_m}(|\omega| z^>) \right] J_{p_m}(|\omega| z^<) \, .
	\label{on-shell propagator}
	\end{align}
This agrees with the result given in Eq.(\ref{G^0 at y=0-result}).


\end{document}